\documentclass{elsart}
\usepackage{graphics}

\newcommand{\antares}{\mbox{ANTARES}}
\newcommand{\daqs}{DAQ system}
\newcommand{\daq}{DAQ}
\newcommand{\adts}{\mbox{all-data-to-shore}}
\newcommand{\rc}{\mbox{RunControl}}
\newcommand{\df}{\mbox{DataFilter}}
\newcommand{\dw}{\mbox{DataWriter}}
\newcommand{\ch}{\mbox{ControlHost}}
\newcommand{\daqh}{\mbox{DaqHarness}}
\newcommand{\timeslice}{time slice}
\newcommand{\on}{on-shore}
\newcommand{\off}{off-shore}
\newcommand{\pmts}{PMTs}
\newcommand{\pmt}{PMT}
\newcommand{\spes}{SPE hits}
\newcommand{\cher}{Cherenkov}
\newcommand{\ars}{ARS}
\newcommand{\floor}{storey}
\newcommand{\floors}{storeys}
\newcommand{\dwdm}{DWDM}
\newcommand{\biolum}{bioluminescence}

\newcommand{\axislabel}[1]      {\textsl{#1}}
\newcommand{\captionfont}[1]    {\textsl{#1}}

\journal{Nuclear Instruments and Methods}

\begin{document}

\begin{frontmatter}

\title{The data acquisition system for the \antares\ neutrino
telescope}

\begin{center}
\renewcommand{\thefootnote}{\fnsymbol{footnote}}
\author[IFIC]{J.A.~Aguilar},
\author[Mulhouse]{A.~Albert},
\author[Rome]{F.~Ameli},
\author[Genova]{M.~Anghinolfi},
\author[Erlangen]{G.~Anton},
\author[Saclay]{S.~Anvar},
\author[CPPM]{E.~Aslanides},
\author[CPPM]{J-J.~Aubert},
\author[Bari]{E.~Barbarito},
\author[LAM]{S.~Basa},
\author[Genova]{M.~Battaglieri},
\author[Bologna]{Y.~Becherini\thanksref{now}},
\author[Bari]{R.~Bellotti},
\author[Saclay]{J.~Beltramelli},
\author[CPPM]{V.~Bertin},
\author[Pisa]{A.~Bigi},
\author[CPPM]{M.~Billault},
\author[Mulhouse]{R.~Blaes},
\author[Saclay]{N. de~Botton},
\author[NIKHEF]{M.C.~Bouwhuis\corauthref{cor}},
\corauth[cor]{Corresponding author}
\ead{mieke.bouwhuis@nikhef.nl}
\author[Leeds]{S.M.~Bradbury},
\author[NIKHEF,UvA]{R.~Bruijn},
\author[CPPM]{J.~Brunner},
\author[Catania]{G.F.~Burgio},
\author[CPPM]{J.~Busto},
\author[Bari]{F.~Cafagna},
\author[CPPM]{L.~Caillat},
\author[CPPM]{A.~Calzas},
\author[Rome]{A.~Capone},
\author[Catania]{L.~Caponetto},
% ...
\nopagebreak[3]
\address[APC]{AstroParticule et Cosmologie, UMR 7164 (CNRS, Universit\'e Paris 7, CEA, Observatoire de Paris), 11, place Marcelin Berthelot, 75005 Paris, France}
\vspace*{-0.40\baselineskip}
\nopagebreak[3]
\address[Bari]{Dipartimento Interateneo di Fisica e Sezione INFN, Via E. Orabona 4, 70126 Bari, Italy}
\vspace*{-0.40\baselineskip}
\nopagebreak[3]
\address[Bologna]{Dipartimento di Fisica dell'Universit\`a e Sezione INFN, Viale Berti Pichat 6/2, 40127 Bologna, Italy}
\vspace*{-0.40\baselineskip}
\nopagebreak[3]
\address[COM]{Centre d'Oc\'eanologie de Marseille, CNRS/INSU Universit\'e de la M\'editerran\'ee Aix-Marseille II, Station Marine d'Endoume-Luminy, Rue de la Batterie des Lions, 13007 Marseille, France}
\vspace*{-0.40\baselineskip}
\nopagebreak[3]
\address[CPPM]{CPPM -- Centre de Physique des Particules de Marseille, CNRS/IN2P3 Universit\'e de la M\'editerran\'ee Aix-Marseille II, 163 Avenue de Luminy, Case 907, 13288 Marseille Cedex 9, France}
\vspace*{-0.40\baselineskip}
\nopagebreak[3]
\address[Catania]{Dipartimento di Fisica ed Astronomia dell'Universit\`a e Sezione INFN, Viale Andrea Doria 6, 95125 Catania, Italy}
\vspace*{-0.40\baselineskip}
\nopagebreak[3]
\address[Erlangen]{Friedrich-Alexander-Universit\"at Erlangen-N\"urnberg, Physikalisches Institut, Erwin-Rommel-Str.\ 1, D-91058 Erlangen, Germany}
\vspace*{-0.40\baselineskip}
\nopagebreak[3]
\address[GEOAZUR]{UMR G\'eoScience Azur, Observatoire Oc\'eanologique de Villefranche, BP48, Port de la Darse, 06235 Villefranche-sur-Mer Cedex, France}
\vspace*{-0.40\baselineskip}
\nopagebreak[3]
\address[Genova]{Dipartimento di Fisica dell'Universit\`a e Sezione INFN, Via Dodecaneso 33, 16146 Genova, Italy}
\vspace*{-0.40\baselineskip}
\nopagebreak[3]
\address[IFIC]{IFIC -- Instituto de F\'{\i}sica Corpuscular, Edificios Investigaci\'on de Paterna, CSIC -- Universitat de Val\`encia, Apdo. de Correos 22085, 46071 Valencia, Spain}
\vspace*{-0.40\baselineskip}
\nopagebreak[3]
\address[IFREMER/Brest]{Centre de Brest, BP 70, 29280 Plouzan\'e, France}
\vspace*{-0.40\baselineskip}
\nopagebreak[3]
\address[IFREMER/Toulon]{Centre de Toulon/La Seyne Sur Mer, Port Br\'egaillon, Chemin Jean-Marie Fritz, 83500, La Seyne sur Mer, France}
\vspace*{-0.40\baselineskip}
\nopagebreak[3]
\address[INAF]{INAF-IASF, via P. Gobetti 101, 40129 Bologna, Italy}
\vspace*{-0.40\baselineskip}
\nopagebreak[3]
\address[IPHC]{IPHC -- Institut Pluridisciplinaire Hubert Curien, UMR 7178 Universit\'e Louis Pasteur Strasbourg 1 and IN2P3/CNRS, 23 rue du Loess, BP 28 -- 67037 Strasbourg CEDEX 2, France}
\vspace*{-0.40\baselineskip}
\nopagebreak[3]
\address[ISS]{Institute for Space Sciences, 77125 Bucharest, Magurele, Romania}
\vspace*{-0.40\baselineskip}
\nopagebreak[3]
\address[ITEP]{ITEP -- Institute for Theoretical and Experimental Physics, B.~Cheremushkinskaya 25, 117259 Moscow, Russia}
\vspace*{-0.40\baselineskip}
\nopagebreak[3]
\address[KVI]{Kernfysisch Versneller Instituut (KVI), University of Groningen, Zernikelaan 25, 9747 AA Groningen, The Netherlands}
\vspace*{-0.40\baselineskip}
\nopagebreak[3]
\address[LAM]{Laboratoire d'Astrophysique de Marseille, CNRS/INSU - Universit\'e de Provence Aix-Marseille I, Traverse du Siphon -- Les Trois Lucs, BP 8, 13012 Marseille Cedex 12, France}
\vspace*{-0.40\baselineskip}
\nopagebreak[3]
\address[LNS]{INFN -- Labaratori Nazionali del Sud (LNS), Via S. Sofia 44, 95123 Catania, Italy}
\vspace*{-0.40\baselineskip}
\nopagebreak[3]
\address[Leeds]{School of Physics \& Astronomy, University of Leeds LS2 9JT, UK}
\vspace*{-0.40\baselineskip}
\nopagebreak[3]
\address[Mulhouse]{GRPHE -- Groupe de Recherche en Physique des Hautes Energies, Universit\'e de Haute Alsace, 61 Rue Albert Camus, 68093 Mulhouse Cedex, France}
\vspace*{-0.40\baselineskip}
\nopagebreak[3]
\address[NIKHEF]{Nationaal Instituut voor Kernfysica en Hoge-Energiefysica (NIKHEF), Kruislaan 409, 1098 SJ Amsterdam, The Netherlands}
\vspace*{-0.40\baselineskip}
\nopagebreak[3]
\address[Pisa]{Dipartimento di Fisica dell'Universit\`a e Sezione INFN, Largo B.~Pontecorvo 3, 56127 Pisa, Italy}
\vspace*{-0.40\baselineskip}
\nopagebreak[3]
\address[Rome]{Dipartimento di Fisica dell'Universit\`a "La Sapienza" e Sezione INFN, P.le Aldo Moro 2, 00185 Roma, Italy}
\vspace*{-0.40\baselineskip}
\nopagebreak[3]
\address[Saclay]{DSM/DAPNIA -- Direction des Sciences de la  Mati\`ere, D\'epartement d'Astrophysique de Physique des Particules de  Physique Nucl\'eaire et de l'Instrumentation Associ\'ee, CEA/Saclay, 91191 Gif-sur-Yvette Cedex, France}
\vspace*{-0.40\baselineskip}
\nopagebreak[3]
\address[Sheffield]{Dept.\ of Physics and Astronomy, University of Sheffield, Sheffield S3 7RH, UK}
\vspace*{-0.40\baselineskip}
\nopagebreak[3]
\address[UU]{Universiteit Utrecht, Faculteit Betawetenschappen, Princetonplein 5, 3584 CC Utrecht, The Netherlands}
\vspace*{-0.40\baselineskip}
\nopagebreak[3]
\address[UvA]{Universiteit van Amsterdam, Instituut voor Hoge-Energiefysica, Kruislaan 409, 1098 SJ Amsterdam, The Netherlands}
\vspace*{-0.40\baselineskip}
\end{center}

\begin{abstract}
The \antares\ neutrino telescope is being constructed in the
Mediterranean Sea.
It consists of a large three-dimensional array of photo-multiplier
tubes. 
The data acquisition system of the detector takes care of the
digitisation of the photo-multiplier tube signals, data transport,
data filtering, and data storage. 
The detector is operated using a control program interfaced with all
elements.
The design and the implementation of the data acquisition system are
described.
\end{abstract}

\end{frontmatter}

\section{Introduction}
\label{sec:introduction}
The \antares\ neutrino telescope~\cite{antares} is used to study
astrophysical sources by detecting the high-energy neutrinos that
these sources may emit.
The detector is deployed on the seabed at a depth of 2.5~km, and
consists of a large three-dimensional array of 900~light-sensitive
photo-multiplier tubes~(\pmts~\cite{pmtpaper},~\cite{ompaper}).
Neutrinos are detected indirectly, after an interaction
inside or in the vicinity of the detector.
The produced charged particles emit \cher\ light, which can be
detected by the \pmts.
From the known positions of the \pmts, and the measured arrival times
of the \cher\ photons, the signal produced by these charged particles
can be distinguished from the background.

The main purpose of the data acquisition (\daq) system
is to convert the analogue signals from the \pmts\ into a format suitable
for the physics analysis.
To achieve this, the \daqs\ has the task to prepare the detector for
data taking, convert the analogue signals from the \pmts\ into digital
data, transport the data to shore, filter the different physics
signals from the background, store the filtered data on disk, and
archive the run settings.
The \daq\ hardware and software are described in
Sections~\ref{sec:daq_hardware} and~\ref{sec:daq_software}
respectively, and the overall performance of the \antares\ \daqs\ is
summarised in Section~\ref{sec:performance}.

\section{\daq\ hardware}
\label{sec:daq_hardware}
\begin{figure}
\begin{picture}(12,18)
\put(0,0){\resizebox{\textwidth}{!}{\includegraphics{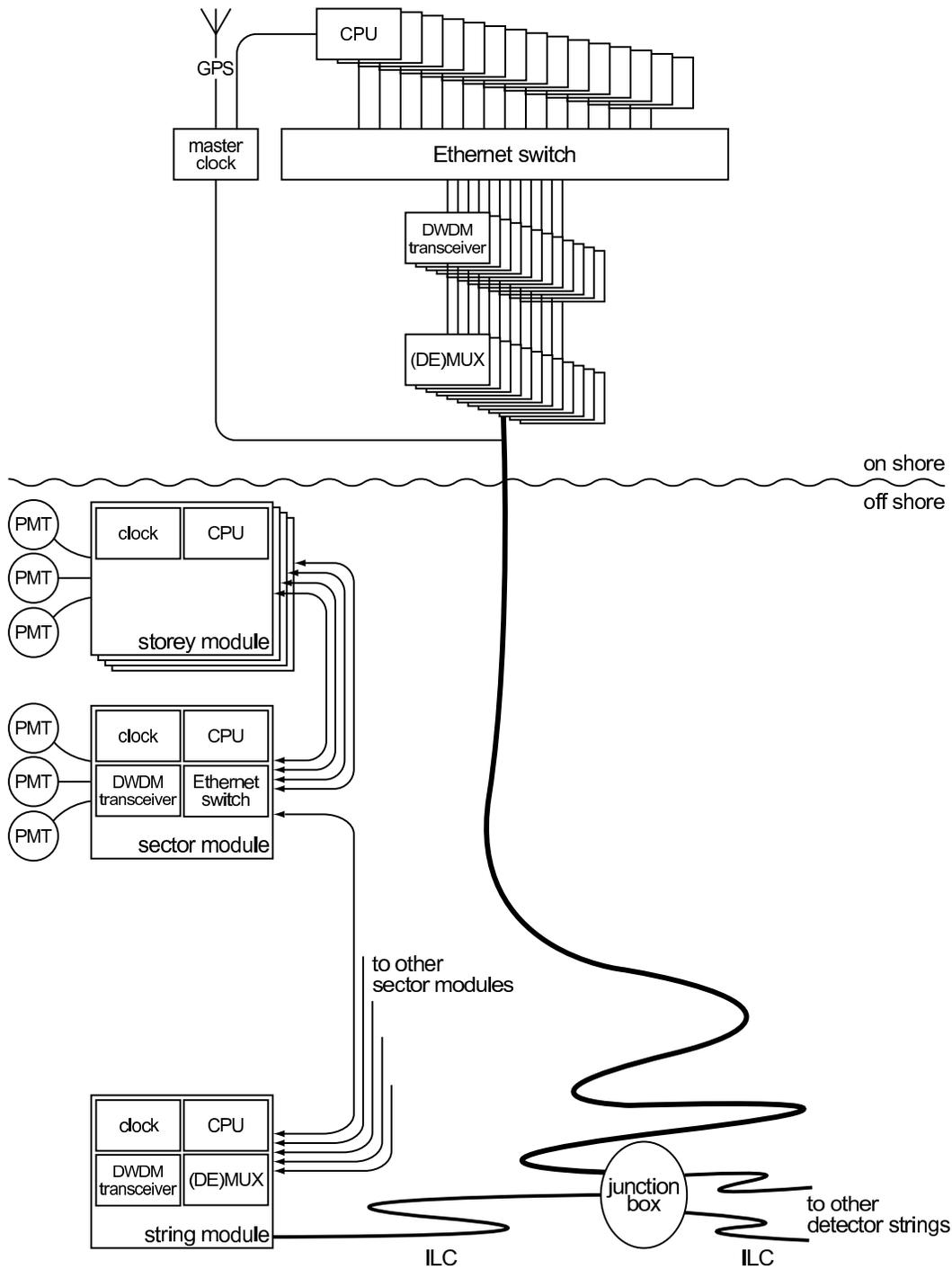}}}
\end{picture}
\caption{\captionfont{Schematic picture of the \daq\ hardware.
The detector consists of 12~vertical strings, each consisting of
25~\floors.
The structure of only one detector string is shown.
Each \floor\ contains three \pmts, and an electronics module with a
local clock and a CPU.
The module on every fifth \floor\ (sector module) contains in addition
an Ethernet switch and a \dwdm\ transceiver (see
Section~\ref{subsec:off_daq_hw}).
Each detector string has at the bottom a string module that contains
an optical (de)multiplexer ((DE)MUX).
Each detector string is connected to the junction box with an
interconnecting link cable~(ILC).
From the junction box a single cable leads to the \on\ PC farm and the
master clock.
The storey, sector, and string module are in \antares\ also referred
to as LCM, MLCM, and SCM respectively.
}}
\label{fig:daq_hardware}
\end{figure}

The hardware of the \daqs\ is a network consisting of hundreds of
processors, as shown schematically in Figure~\ref{fig:daq_hardware}.
A significant part of this network is located off shore.
The \off\ processors are integrated in custom made electronics.
The \off\ part of the \daqs\ is connected to the \on\ part by a single
electro-optical cable.
The \on\ processors are standard PCs and are located in the shore
station.

\subsection{Off-shore \daq\ hardware}
\label{subsec:off_daq_hw}
\begin{figure}[h]
\begin{picture}(12,8)
\put(0,0){\resizebox{\textwidth}{!}{\includegraphics{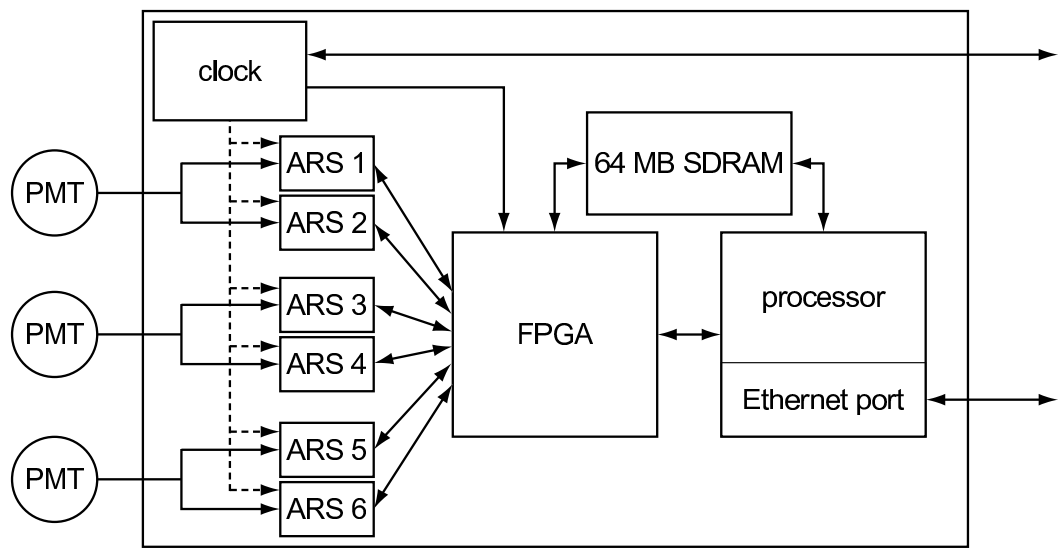}}}
\end{picture}
\caption{\captionfont{Schematic picture of the main hardware
components in the electronics module of a \floor.}}
\label{fig:electronics}
\end{figure}

Most of the \off\ electronics modules are used to read out the
photo-multiplier tubes (\pmts).
These modules are indicated in Figure~\ref{fig:daq_hardware} by `sector
module' and `\floor\ module'.
A detector string is divided into 25~\floors, each consisting of a
triplet of \pmts\ and a titanium container.
This container houses the electronics needed for the
digitisation of the analogue 
signals from the \pmts, and the data transport. 
The hardware components that play a key role in the
\daqs\ are shown schematically in Figure~\ref{fig:electronics}.
These components include several custom designed analogue ring sampler
(\ars) chips~\cite{ars}, a local clock, a field programmable gate array
(FPGA), an SDRAM of 64~MB, and a processor with an Ethernet port. 

During data taking the analogue signals from the \pmts\ are processed
by the \ars\ chips.
This includes the digitisation of the timing of each \pmt\
signal and the total charge of the pulse.
The settings of each \ars\ chip can be adjusted, including the two main
parameters: threshold and integration gate.
The voltage threshold is set to eliminate small pulses due to the dark
current in the \pmt. 
Its typical value corresponds to 0.3~photo-electrons.
The gate is set to integrate most of the \pmt\ signal, while
limiting the contribution of electronic noise.
The typical width of the integration gate is 35~ns.
A local clock is used by the \ars\ chips to timestamp each \pmt\ signal
above threshold.
The combined time and charge information of the \pmt\ signal within
the integration time is referred to as a single photo-electron hit.
After the integration time, the \ars\ chip has a dead time of about
200~ns due to the limited transfer speed to the analogue pipeline.
To compensate for this dead time, two \ars\ chips are connected to each
\pmt.
These \ars\ chips operate in a token ring scheme.
After the integration time of the first \ars\ chip, the second 
takes over.
Only after the integration time of the second \ars\ chip, the dead time
of the first plays a role.

For special cases, like large or double pulses, the pulse
shape can be read out by the \ars\ chip with a sampling
frequency that is tunable between 150~MHz and 1~GHz.
This feature is particularly useful for calibration and tuning of
detector parameters, but can also be enabled for physics triggers.
The pulse shape information is not used in the data filter algorithms
(see Section~\ref{subsec:data_taking}), but it can be stored on disk with
the associated hit information.

The readout system of the \ars\ chips is implemented in a high
density FPGA XilinX Virtex-E XCV1000E.
The data from the \ars\ chips are buffered by the FPGA in the 64~MB
SDRAM into separate frames covering a certain period.
This period can be set to values between 10~and 100~ms, and
corresponds to a predefined number of clock cycles. 
The clock system is described in more detail in
Section~\ref{subsec:clock}.

The processor in the \off\ electronics modules is the
interface between the \ars\ chips and the online data processing
system.
It is the Motorola MPC860P, which has a fast Ethernet port~(100~Mb/s).
The operating system on these processors is
VxWorks\footnote{http://www.windriver.com/}.
TCP/IP is used for the data transport.

The module on every fifth \floor\ on a detector string, in
Figure~\ref{fig:daq_hardware} specified by `sector module', has some
additional functionality.
These modules contain an Ethernet switch, and the transceiver that
plays a role in the data transport to and from shore.
The Ethernet port on each \floor\ is connected to a sector module
using an optical bidirectional 100~Mb/s link.
The links from five \floors\ are merged by the Ethernet switch in the
sector module into a single Gb/s Ethernet link.
The group of \floors\ with a common Gb/s Ethernet link is referred to
as a sector.
For a total of 25~\floors\ on a detector string, each detector string
has five such sectors.

Each detector string has at the bottom an extra container which houses
an electronics module, in Figure~\ref{fig:daq_hardware} specified by
`string module'.
The string module is used for the slow control of the
electrical power system and the calibration systems of the detector
string, and also for the distribution of the clock signal.
This module has a separate 100~Mb/s link to shore.

For the data transport between the \off\ and \on\ parts of
the detector the dense wavelength division multiplexing (\dwdm)
technique is used~\cite{dwdm}.
This is an optical technology that uses multiple wavelengths to
transmit different streams of data along a single fibre.
The \dwdm\ system can be considered as a large fibre-optic network.
The \dwdm\ transceivers are used to extend the Ethernet link to shore.
The five \dwdm\ channels from (to) the sectors and the \dwdm\ channel from
(to) the string module are optically (de)multiplexed in the string
module.
Each sector and each string module has a unique pair of wavelengths
that is used to transmit the data.

Each string is connected with an electro-optical cable to
the so-called junction box, the container where the cables from the
12~detector strings meet.
The junction box is connected to the shore station with a single 40~km
long electro-optical cable. 
On shore, there is a multiplexer and a demultiplexer for each detector
string.
The \dwdm\ system is also used for the transmission of the slow control
data, and the distribution of initialisation and configuration data.
The data that are sent from the shore station to the detector are
multiplexed on shore, and demultiplexed in the string modules.

\subsection{On-shore \daq\ hardware}
\label{subsec:on_daqhw}
The \on\ part of the \daqs\ is located in the shore station.
It consists of a farm of standard PCs, a large Ethernet switch, the
\dwdm\ hardware, and the master clock system (described in
Section~\ref{subsec:clock}).

All PCs in the farm have software programs running that are needed for
the detector operation, data transfer and communication, data
processing, monitoring, and data storage.
These programs are all part of the \daq\ software, discussed in
Section~\ref{sec:daq_software}. 
The PC farm consists of some tens of PCs, all connected to the
Ethernet switch.
With 12~detector strings, each consisting of 25~\floors\ and an
electronics module at the bottom, there are
312~\off\ processors, each with an Ethernet port for the
connection to shore.
All \on\ and \off\ processors that are part of the \daqs\ are
connected to the same \on\ Ethernet switch.
Together they form a large Ethernet network.
Each of these processors can be considered as a node in the \daq\
network, addressable by its unique IP~address.
All \off\ processors can be reached by a telnet connection.
This allows system tests to be performed and new software and
firmware to be downloaded.
The Ethernet network makes the distribution of the data from the \ars\
chips to the \on\ processes completely transparent, and it enables the
communication with all processors in the system.

Like in every string module, a (de)multiplexer and \dwdm\ transceiver
is available on shore for each detector string.
The multiplexers are used for multiplexing the data streams that
are meant for the initialisation and configuration of the detector,
and the demultiplexers are used for demultiplexing the data streams
from the \ars\ chips, and slow control data for monitoring purposes.

\subsection{Clock system}
\label{subsec:clock}
The main purpose of the clock system is to provide a common clock
signal to all \ars\ chips.
It consists of a clock signal generator on shore, a clock signal
distribution system, and a clock signal transceiver in each detector
\floor.
The \on\ master clock generates a 20~MHz clock signal that is
synchronised internally to the GPS time with an absolute accuracy of
100~ns.
The clock signal is distributed to all \off\ clock transceivers using
the available optical fibre network.
Therefore each detector \floor\ has one local clock that is
synchronised to the master clock.
The phase offset of each local clock can be determined by measuring
the return time of a calibration signal.
The local clock is used to synchronise the \ars\ chips.
The relative time accuracy is about 50~ps.
The clock system operates in parallel and independent of the \daqs.
As a result, the internal clock calibration does not induce any
dead time.

\section{\daq\ software}
\label{sec:daq_software}
\begin{figure}[b]
\begin{picture}(12,10)
\put(0,0){\resizebox{\textwidth}{!}{\includegraphics{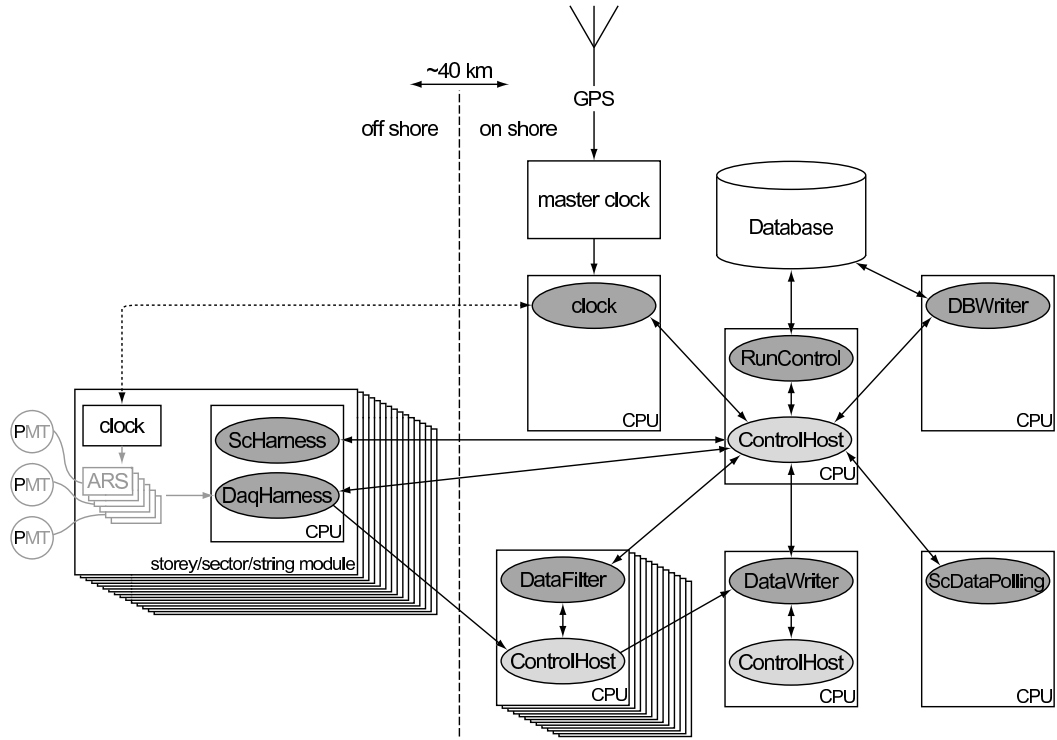}}}
\end{picture}
\caption{\captionfont{Schematic overview of the main processes in the
\daqs. 
The solid lines indicate the links between processes, for which TCP/IP
is used. 
The dotted line indicates the distribution of the reference
clock signal. 
The processors in each \off\ electronics module, whether this module
is connected to \pmts\ or not, has the same processes running.
See text for an explanation of the processes.}}
\label{fig:daq_software}
\end{figure}

The main software processes in the \daqs\ are indicated schematically
in Figure~\ref{fig:daq_software}.
Part of the processes run on the \off\ processors, part of them run on
the \on\ processors.
There are basically three types of processes: processes that are used
for the data transfer and communication (see
Section~\ref{subsec:data_transfer}), processes that take care
of the operation of the detector (see
Section~\ref{subsec:detector_operation}), and processes that are involved
in the data taking and data handling (see
\mbox{Sections~\ref{subsec:detector_operation}--\ref{subsec:data_storage}}).
In total, with hundreds of \off\ processors and the \on\ PC farm, there
will be several hundreds of processes in the \daqs.
A \daq\ model, that is implemented in each process, has been defined to
synchronise all processes.

\subsection{\daq\ model}
\label{subsec:daq_model}
\begin{figure}[h]
\begin{picture}(12,3.8)
\put(0,0){\resizebox{\textwidth}{!}{\includegraphics{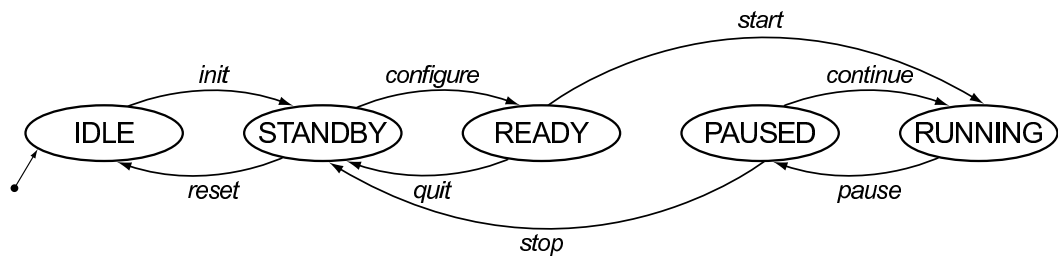}}}
\end{picture}
\caption{\captionfont{The finite state machine of the \daqs. 
The ellipses indicate the different states, the arrows indicate the
state transitions.
When the state machine is entered it starts in the \textsc{idle} state.}}
\label{fig:state_m}
\end{figure}

The hundreds of processes in the distributed multiprocessor system
operate independently.
In order to synchronise all processes in the system, a finite state
machine has been designed with a fixed set of states and transitions,
represented by the state diagram shown in Figure~\ref{fig:state_m}.
Each of the processes that participates in the \daqs, except for the
data transfer package \ch\ (see Section~\ref{subsec:data_transfer}),
has this state machine implemented, for which the CHSM
language~\cite{chsm} is used.
The state transitions are accompanied by actions
that are specific to each type of process in the system.
These actions should be performed before the data taking can start and
after the data taking has stopped.
The state machine ensures that these actions are executed for all
processes in the required order (as specified by the state
transitions).

The steps involved in the preparation of the \daqs\ for data taking
include the configuration of the hardware components and software
processes.
The required steps for stopping the data taking (and switching off the
detector) include the handling of the data that are still being
processed, closing the data files, and storing the run information in
the database.
Data are only taken in state \textsc{running}.
Each start transition corresponds to a new data taking run, identified
by a unique run number.
The run number is added to the raw data by the \off\ processes, and is
also used by the data storage program to archive the data.

Processes in the \daqs\ can only be in one of the predefined states.
State transitions are initiated from a central application
(see Section~\ref{subsec:detector_operation}), and apply to all
processes in the system.
During a state transition, all processes perform the necessary actions
independently.
Consequently, the actual state of the system is associated with all
actions that took place during the state transitions.
Normally all processes are in the same state at any time.

\subsection{Data transfer and communication}
\label{subsec:data_transfer}
All data transfer and communication between processes in the \daqs\ are
done with the \ch\ package~\cite{controlhost}.
\ch\ implements the tagged data concept, which means that any program in
the system can send data accompanied by a tag to a \ch\ server.
The server will distribute the data to all processes that subscribed
to the specific tag using TCP sockets.
Most of the PCs in the shore station have a \ch\ server running.
The \ch\ package has been optimised for Gb/s Ethernet, and its
overhead is typically 5\% of the total bandwidth.

\ch\ is used for the data transfer between processes at three
different stages in the \daq: data processing, data storage, and
detector operation.
The data from the \off\ processes (\daqh) are transferred to the
\on\ processes that take care of the processing of the data
(\df, see Section~\ref{subsec:data_taking}).
On each PC in the shore station that has such a data processing
program running, a \ch\ server is running for passing the data.
All data that passed the data processing, and that have to be stored
on disk, are transferred from the data processing PCs via one \ch\
server to the data storage program (\dw, see
Section~\ref{subsec:data_storage}).
The \rc\ program, that is used to operate the detector
(see Section~\ref{subsec:detector_operation}), communicates with all
processes in the \daqs.
All state transitions
are initiated by the \rc.
The commands for the state transitions,
as well as the messages for the handshaking mechanism are
distributed via a central \ch\ server.

\ch\ takes care of the distribution of a single message to multiple
processes on multiple hosts.
Without knowing how many processes should receive a certain message,
each will receive the message if it subscribed to the corresponding
tag.
The implementation of \ch\ also enables the exchange of
information between processes written in different programming
languages ($\rm C^{++}$ and Java{\tiny\texttrademark} are
used in the \daqs). 
The \rc\ program is written in Java{\tiny\texttrademark}
because of the straightforward possibility for remote operation of the
detector, the user friendly interface with the
Oracle$^{\tiny\textregistered}$ database that is used in the \daqs,
and the advanced toolkit for developing graphical user interfaces. 
The \on\ programs used for data processing and data handling are
written in $\rm C^{++}$ because of the required execution speed.
The required compatibility of the \off\ software with the related
firmware also implies the use of the $\rm C^{++}$ programming language.

\subsection{Detector operation}
\label{subsec:detector_operation}
The main program in the \daqs\ that is used to operate the detector is
the \rc.
It has a graphical user interface from which the operator submits the
requests for the execution of state transitions by all processes in
the system (shown in Figure~\ref{fig:state_m}).
All processes that are subjected to these state transitions can be
considered as client processes.
With these state transitions the detector can be prepared for data
taking, and the necessary actions are performed when the data taking
has ended.

All data needed by the \daqs\ to operate the detector are stored in
the central database system, for which the relational database system
Oracle$^{\tiny\textregistered}$ is used.
These data are retrieved from the database by the \rc\ only.
These data include
the detector information (the location of each component
in the detector), the client process list, the different configurations
for each client setup, the initialisation and configuration
data, and the information about the data taking.
The client process list determines which processes participate in the
system, and thus which parts of the detector are active
during data taking.
The \rc\ retrieves the necessary information from the database by
using the input from the operator and the relations between
the database tables.

A separate graphical user interface is used to control the power
system in each string module.
When the \off\ electronics modules are powered on, the \off\
processors boot.
The \rc\ launches each client process that is part of the client setup
chosen by the operator by executing its start command on the host with
its associated IP~address (both retrieved from the database).
At startup the client processes enter the general state machine.

Once the client processes are started, the general mechanism that is
used for the communication between the \rc\ and its clients is done
with a handshaking method based on the unique~ID of each
process.
This ID consists of the nickname of the process (which corresponds to
the client type), and the IP~address of its host.
The implementation of the handshaking method is that, after the
request for a state change from the \rc\ via the central \ch\
server, each client sends a message containing this unique~ID, and the
state it just entered, to the
same \ch\ server that indicates that the state change has
succeeded for this client. 
As the \rc\ uses the same unique coding for each client, it is
aware of the current state of each client.
As a consequence, the \rc\ also has a monitoring task,
visualising the state of each client. 
The \ch\ package is designed in such a way that it broadcasts a
message when a process connects to or disconnects from a \ch\ server.
These messages are also intercepted by the \rc.
As each client connects to the \ch\ server when it is started, and
disconnects when it terminates, the \rc\ also
visualises the running state of each client.
In addition, the \rc\ provides the means to modify the client
setup during operation in case of a client error to prevent
interruption of the data taking.

Depending on the type of data that will be taken, a predefined
configuration for the chosen client setup is selected by the
operator.
The specific initialisation and configuration data required by each
process are retrieved from the database by 
the \rc, and sent, together with the corresponding state
transition request, via the \ch\ server to the particular client
process.
Along with the state transitions for starting and stopping a run,
the master clock on shore sends a hardware signal to the \off\ clocks
for the synchronisation of the \ars\ chips. 

The ScHarness process that runs on each \off\ processor controls
the high voltage
applied to the \pmts\ during the initialisation and configuration
phase. 
It is also used for the readout of various instruments.
The data from these instruments are sent to shore, and stored in the
database by the \mbox{DBWriter} program.
These data are then used to monitor the detector and its environment.
The ScDataPolling program schedules the read requests of all these
instruments in the detector.

The \rc\ updates the database with the run information and the
reference to the detector settings to be able to couple the detector
settings to the data when they are eventually analysed.
A new run is automatically started after a few hours of data taking,
or when the output data file, created and updated by the data storage
program (see Section~\ref{subsec:data_storage}), reaches its maximum
size (typically 1~GB).
The \rc\ then triggers the necessary state transitions automatically
to stop the run, and start a new run.

During data taking, the raw data are monitored
online using dedicated software.
For this, the data are taken from \ch\ and histogrammed using the
ROOT package~\cite{root}.
A custom made presentation tool based on ROOT displays and compares 
histograms, showing the basic measurements such as the time and
amplitude information from the \pmts.

\subsection{Data taking and data processing}
\label{subsec:data_taking}
During data taking, all signals that are recorded by the \pmts\ and
digitised off shore are transported to shore without any \off\
data selection.
This implementation is known as the `\adts' concept.
As a result, all raw data are available on shore where the required
data processing method can be applied to the data.
This minimises the overall loss of physics signal, and allows
different processing methods to be applied for specific physics
analyses.

The \on\ handling of all raw data is the main challenge in the
\antares\ \daqs, because of the high background rates,
mainly caused by \biolum\ and $^{40}$K decay.
As a result of this background, the average photon detection rate
per \pmt\ varies between 60~kHz and 90~kHz during periods with low
\biolum\ activity~\cite{background}.
The data output rate of the detector is then 0.3~GB/s to 0.5~GB/s.
The \on\ handling of the data is done with the \df\ processes that
run on each PC in the \on\ data processing farm.
The processing of the data involves the separation of 
signal from background, and as a result the necessary reduction of
the data flow for storage.
During data taking, the \df\ programs process all data online as the
detector is in principle operating continuously.
The \df\ programs apply various algorithms, each aimed at
the search for specific physics signals.
This includes a muon filter that looks in all directions for the
signals in the detector compatible with a muon.
Depending on the random background rate, the filter settings can be
adjusted so that the trigger rate is dominated by muons passing the
sensitive volume of the detector.
For an average data size of a physics signal of 2.5~kB, and an
expected atmospheric muon rate of about 10~Hz, the data reduction is
then about~$10^5$. 
The bandwidth required for the data transport scales linearly with the
background photon detection rate.
The data processing speed scales with some power of this rate,
this power being the required minimum number of photons to identify a
physics signal (see Section~\ref{sec:performance}). 

\begin{figure}
\resizebox{\textwidth}{!}{\includegraphics{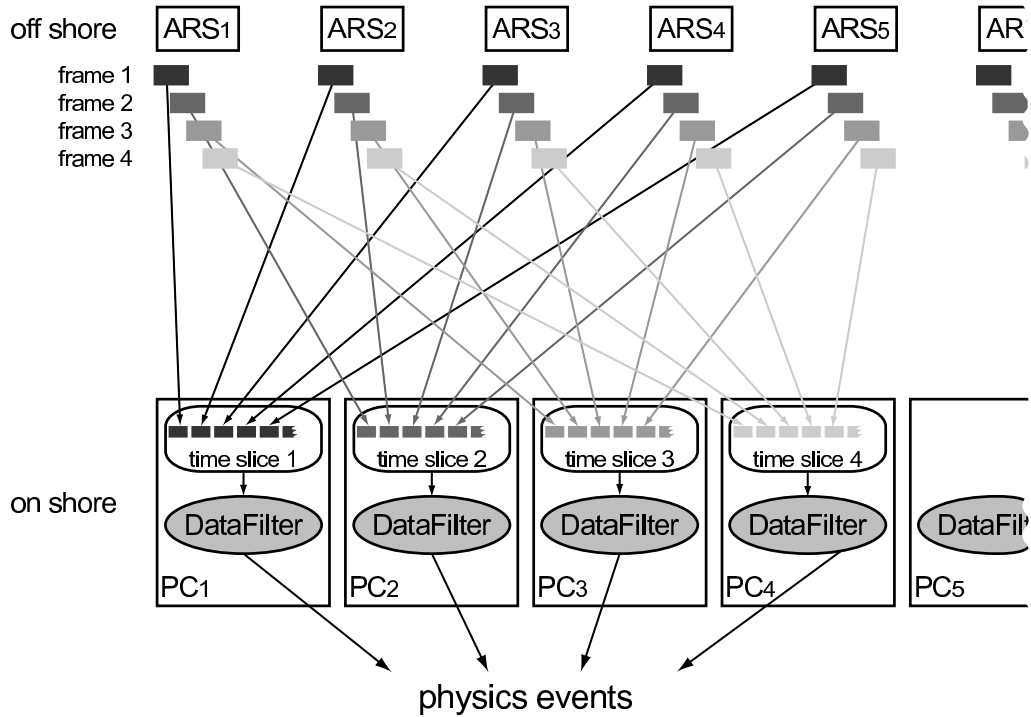}}
\caption{\captionfont{The processing of the data based on \timeslice s.
All frames belonging to the same time window are sent to a single PC
and form a \timeslice.
The \df\ program running on each PC processes the data in the
\timeslice.
All physics events are stored on disk (see
Section~\ref{subsec:data_storage}).
}}
\label{fig:timeslices}
\end{figure}

The \df\ processes receive the raw data from the \off\
\daqh\ processes, running on each \off\ processor in the detector.
All data that are produced by each \ars\ chip in a certain time window
are buffered in the SDRAM by the FPGA in a so-called frame.
The length of this time window can be set to values between 10~and
100~ms.
A separate thread that is spawned by the \daqh\ process takes the
buffered data from the SDRAM, and sends each
frame as a single packet to shore using TCP/IP (see
Figure~\ref{fig:timeslices}).
The frames from all \daqh\ processes that belong to the same time
window are sent to the same PC in the \on\ data processing system,
identified by its IP~address.
The large Ethernet switch thus operates as a so-called level~2 switch.
The list of IP~addresses as possible destinations for the raw data
is provided to the \daqh\ processes by the \rc.
The collection of frames belonging to the same time window is called
a \timeslice.
As a result, a \timeslice\ contains all data that were registered
during the same time interval by all \ars\ chips in the detector.
The \df\ programs alternately receive the frames belonging to
the same time window from the \off\ processes via the \ch\ server that
runs on the same PC.
Each \df\ program has to be finished with processing a \timeslice\
before it receives the next.
This imposes an optimisation of the configuration of the \df\ programs
in terms of processing speed, and it determines the number of PCs
required for online data processing and the specifications of these
PCs.

When a sufficient number of correlated single photo-electron (SPE)
hits is found that is consistent with a 
specific physics signal, the data are
considered as a physics event, that is written to disk.
The period covered by a \timeslice\ is chosen such that its duration
is long compared to the duration of a physics event in the detector
(approximately 1~$\mu$s), in order to
minimise the chance of having an event crossing the boundaries of a
\timeslice.

The algorithms implemented in the data processing software are
designed to find a physics signal by looking for
space-time correlations in the data.
From the time of the \spes,
and the positions of the \pmts, these algorithms calculate in real
time if hits could originate
from light produced by this signal.
In the case of standard data processing, when the muon filter is used,
this is the signal produced by a muon traversing the detector.
Apart from the standard muon filter there are various other algorithms
implemented that are used for the search for specific physics signals.
The configuration of the \df\ programs, that indicates which algorithm
is used, is done by the \rc, as described in
Section~\ref{subsec:detector_operation}.
Among the presently available filter algorithms is a magnetic monopole
filter, an optical beacon filter, and a directional filter.
The monopole filter looks for space-time correlations in the data that
are caused by particles that traverse the detector at lower speeds
than a muon.
The optical beacon filter is used when the optical beacons, that are
attached to the detector strings, are fired.
These optical beacons produce a typical light signal, that is used for
time calibration.
The directional filter is used when searching for a muon coming from a
specific direction.
This filter method takes into account the known direction of
the neutrinos, and is used for astrophysical sources with known
positions on the sky.
For gamma-ray bursts, a subgroup of these sources, this filter method
can be used offline.
The combination of the \adts\ concept, the use of an \on\ PC farm, the
specific features of gamma-ray bursts, and the information provided by
external satellite triggers, make the \antares\ detector particularly
sensitive to neutrinos from gamma-ray bursts~\cite{grb}.

Any other possible filter algorithm for a specific
physics signal can easily be implemented in the data processing
system.
As the whole data processing system is implemented in software, a new
filter algorithm only involves adding the specific code and 
the corresponding configuration option in the database.
If necessary, the computing power can be increased by adding PCs to
the \on\ system.

\subsection{Data storage}
\label{subsec:data_storage}
The physics events selected by the \df\ programs are passed via one
\ch\ server to the \dw, which formats the events and writes them to
disk in ROOT format~\cite{root}.
These disks are located in the shore station.
For each run, the data are stored in a separate file.
The event files are used for the physics
analyses, and are copied regularly from the shore station to a
computer centre elsewhere.

\section{Performance of the \daqs}
\label{sec:performance}
The performance of the \daqs\ has been evaluated on the basis of
the operation of a prototype detector~\cite{background}, and has been
quantified using a detailed simulation of the response of the complete
detector to muons~\cite{MC}.
During periods of low \biolum\ activity (when the average photon
detection rate is 60--90~kHz per \pmt) all data are
transmitted from the \off\ electronics modules to the shore station.
The total data output rate of the complete detector is then
0.3--0.5~GB/s.
However, at higher background rates congestion of the data flow can
occur at the processors in the \off\ \floor\ modules.
Although these \off\ processors have a 100~Mb/s Ethernet port, they
cannot make use of the full available bandwidth.
The maximum data throughput of a single \floor\ is at present 20~Mb/s.
This corresponds to a photon detection rate of about 150~kHz per
\pmt.
With the zero-copy feature of the VxWorks operating system running on
the \off\ processors, this maximum data flow could be increased to
about 40~Mb/s.
This corresponds to a maximum manageable photon detection rate of
about 300~kHz per \pmt.

The sector modules have a Gb/s Ethernet link to shore which can easily
accommodate the maximum rate.
On shore, there is a single large Ethernet switch to which all data
processing PCs, each with a Gb/s Ethernet link, are connected.
To avoid congestion of the data flow at the data processing PCs, a
barrel shifting mechanism of the data has been implemented in the
\off\ \daqh\ processes.
In this mechanism, the data sending threads do not send the data for a
given \timeslice\ simultaneously to the same \on\ PC, but
desynchronise the data sending according to a predefined scheme.

\begin{figure}[t]
\begin{center}
\begin{picture}(10,9)
        \put(0,0){\scalebox{1}{
                \put(0.6,0.5){\scalebox{0.6}{\includegraphics{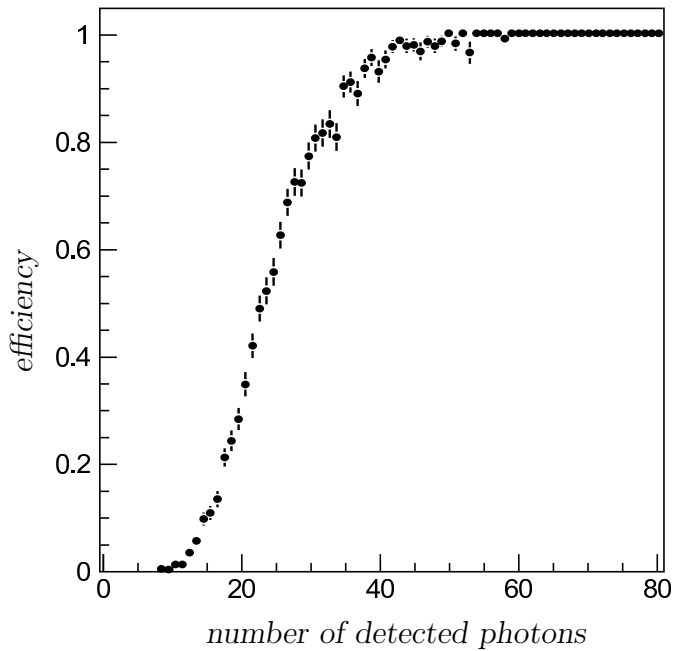}}}
                \put(0,3.6){\rotatebox{90}{\axislabel{efficiency}}}
                \put(2.6,0){\axislabel{number of detected photons}}
        }}
\end{picture}
\end{center}
\caption{\captionfont{Efficiency of the standard muon filter as a
function of the number of detected photons for a given event, using a
simulation of the detector response to muons. 
The efficiency is defined as the ratio between the number of events
found by the filter and the number of generated events.
}}
\label{fig:efficiency}
\end{figure}

The filter programs that run on the data processing PCs have to filter
the data in real time.
The number of required filter programs, which corresponds to the
number of data processing PCs involved in the \daqs, depends mainly on
the background rate in the sea.
The standard muon filter is configured such that it triggers when it
finds 5 or more space-time correlated level~1 hits (two hits within
20~ns in one detector \floor, or one hit with an amplitude larger than
2.5~photo-electrons).
Therefore a minimum of 10 detected photons is required to trigger an
event.
The trigger efficiency of this filter has been evaluated using a
simulation of the detector response to muons~\cite{MC}, and the result
is shown in Figure~\ref{fig:efficiency} as a function of the number of
detected photons for a given event.
The efficiency rises rapidly above 10 detected photons,
and reaches~1 at 40 detected photons.

During periods of low \biolum\ activity (see above), and using a
trigger based on a majority logic, the trigger rate
due to random background is a few hundred~Hz.
Simulations have shown that with a more sophisticated trigger, this
rate can be reduced to less than 1~Hz, without loss of efficiency.
With an expected atmospheric muon trigger rate of about 10~Hz, the
standard muon filter then has a purity of 90\%.
With the filter conditions mentioned above, 10~PCs with a processing
speed of 3~GHz are needed.
Different filter configurations are used for the various physics
filters described in Section~\ref{subsec:data_taking}.
For physics filters that require more computing power, or that use
less strict filter conditions (as used for example in the directional
filter), more data processing PCs are used to be able to filter the
data online.

In principle the data taking and online data processing take place
continuously.
There is no dead time in the data transport, nor in the
online filtering of the data.
Only a small dead time is encountered in the \ars\ chips, such that a
third photon is lost when it is detected by one \pmt\ within 235~ns
(i.e. integration gate and dead time of the \ars\ chip).
The effective dead time is then less than 165~ns, taking into account
the integration gate of both \ars\ chips.
However, at high background photon detection rates, the maximum
bandwidth of the \off\ processors is reached (see above).
Part of the data is then lost, and this will result in a loss of the
overall detection efficiency.
With the prototype detectors and the first complete detector string
that have been operational, significantly higher background
rates than the maximum manageable background rate have been measured,
especially during spring.

\section{Conclusions}
\label{sec:conclusions}
The \antares\ \daqs\ has been operational since~2003, and has been
used for various systems that have been deployed in the deep-sea
environment.
The \adts\ concept offers a flexible data filtering system.
A simulation of the detector response to muons has shown that a high
detection efficiency can be achieved.
The results obtained with one of the prototype detector strings have
been published~\cite{background}.
For this detector, the
data taking was more than 90\% efficient during periods of low
\biolum\ activity.
The \daqs\ has also been used to read out the first
complete detector string.
By enlarging the \on\ computer farm, the same system will be used to
read out the complete detector, that will consist of twelve
detector strings.

\section*{Acknowledgements}
The authors would like to thank R. Gurin and A. Maslennikov for
providing the \ch\ package.
The authors also acknowledge the financial support of the funding
agencies:
Centre National de la Recherche Scientifique (CNRS),
Commissariat \`a l'Energie Atomique (CEA),
Commission Europ\'eenne (FEDER fund),
R\'egion Alsace (contrat CPER),
R\'egion Provence-Alpes-C\^ote d'Azur,
D\'epartement du Var and Ville de La Seyne-sur-Mer,
in France;
Bundesministerium f\"ur Bildung und Forschung (BBF),
in Germany;
Istituto Nazionale di Fisica Nucleare (INFN),
in Italy;
Stichting voor Fundamenteel Onderzoek der Materie (FOM),
Nederlandse organisatie voor Wetenschappelijk Onderzoek (NWO),
in The Netherlands;
Russian Foundation for Basic Research (RFBR),
in Russia;
Ministerio de Educaci\'on y Ciencia (MEC),
in Spain.

\end{document}